\documentstyle[aps,prb,preprint]{revtex}

\begin {document}

 \title {Geometrical-confinement effects on excitons in quantum disks}

 \author{Jakyoung Song and Sergio E. Ulloa}

 \address{Department of Physics and Astronomy and Condensed Matter and
  Surface Sciences Program \\ Ohio University, Athens, Ohio 45701--2979}

  \maketitle

 \begin{abstract} 
 
Excitons confined to flat semiconductor quantum dots with elliptical
cross section are considered as we study geometrical effects on exciton
binding energy, electron-hole separation, and the resulting linear
optical properties.  We use numerical matrix diagonalization techniques
with appropriately large and optimized basis sets in an effective-mass
Hamiltonian approach.  The linear optical susceptibilities of GaAs and
InAs dots for several different size ratios are discussed and compared
to experimental photoluminescence spectra obtained on 
GaAs/Al$_x$Ga$_{1-x}$As and 
InAs/GaAs quantum dots.  For quantum dots with several nm in size, 
there is a strong blue shift of the luminescence due to geometrical
confinement effects.  Also, transition peaks are split and shifted
towards higher energy, in comparison with dots with circular cross
sections.
 
 \end{abstract} 

 
 \narrowtext

 \section{ Introduction }

The development of clever fabrication techniques in semiconductors has
brought the reduction of the effective dimension of electronic states
from their usual three-dimensional character in bulk materials, to 
``zero-dimensional'' states in quantum dots.\cite{1}  The quantum effects
of these lower-dimensional systems have attracted much attention in
recent years, due in part to possible applications which include
electronic devices based on parallel and perpendicular transport,
quantum well lasers, and optical devices. \cite{2}  Two-dimensional
quantum-well or quantum-film structures, which provide confinement in
one space dimension, have been well investigated, and quantum exciton
effects observable even at room temperature have been studied.\cite{3} 
The confinement of excitons has also been shown to result in very large
electro-optical shifts of the absorption peaks, producing the so-called
quantum-confined Stark effect.\cite{3,4}
                                            
In quasi-zero-dimensional quantum dot systems, the additional  quantum
confinement dramatically changes the optical and electronic properties,
compared to those in higher-dimensional structures, as the whole
single-particle spectrum is now discrete.  Correspondingly, the
excitonic spectrum is expected to be strongly affected.  The properties
of excitons confined in quantum boxes were first analyzed theoretically
by Bryant,\cite{5} who used variational and configuration-interaction
representations.  Later on, excitons and biexcitons have been studied,
\cite{6,7} as well as excitons in the presence of a strong magnetic
field,\cite{8} using numerical matrix diagonalization schemes.   

On the experimental side, interband optical spectroscopies, such as
photoluminescence, have been used to study various quantum dot systems
--- such as those produced in the GaAs/Al$_x$Ga$_{1-x}$As structure, with its
bandgap modulation.\cite{9,10,11}  More recently, fascinating studies
on so-called ``self-assembled'' quantum dots, such as 
InAs and In$_x$Ga$_{1-x}$As
clusters on GaAs substrates, have also been reported.\cite{12,13,14,15}
 Most of the theoretical investigations are based on the assumption
that the shape of quantum dots is a simple sphere or box, having a
great deal of symmetry, both because  it simplifies calculations and
because quantities such as the exciton binding energy scale very well
with the overall dot size.   However, realistic dot shapes are probably
much less symmetrical, as well as being typically flat and more
two-dimensional in shape.\cite{12,13,14,15}  
                                           
Here, we consider the effect that less symmetric structures, namely
flat quantum dots with elliptical cross sections, or ``elliptical
quantum disks'', have on the excitonic optical properties. To date,
little work has been reported on the properties of nonsymmetric
quantum dots, probably because this system has more complicated
solutions.\cite{13,15}  Our studies within the effective mass
approximation yield some very interesting consequences of the
elliptical asymmetry: apart from the expected blue shift of the first
excitonic transition for dots with the same overall area but different
axes, we find a rearrangement of the oscillator strength which
characterizes individual dot shapes.  In particular, since elliptical
cross-section dots have less symmetry, some of the accidental
degeneracies in circular dots giving rise to stronger and fewer peaks
in the imaginary part of the optical susceptibility are split.  This
gives rise to a more monotonically decreasing peak intensity for higher
energy features in the susceptibility of noncircular dots.  This
behavior can in turn be used to structurally characterize specific dots
from their photoluminescence excitation response.  
                                   
The remainder of the paper is organized as follows.   We introduce the
theoretical method in Sec. II.   Here, we outline the effective mass
Hamiltonian approach and introduce the various basis function
representations which allow us to use numerical methods to calculate
the eigenvalues and  eigenfunctions of excitons in these quantum dots. 
A great deal of care is needed to assure that the solutions obtained
are well behaved and converged with a finite computational effort.  We
discuss in this section how this is accomplished.  In Sec. III, we
discuss the main geometrical effects on various exciton
characteristics, such as the exciton binding energy, electron-hole
separation, and the linear optical susceptibility.  Solutions for
excitons in quantum dots with both circular and elliptical
cross sections are shown, using large enough basis sets, and a set of
optimized basis functions, which improve the accuracy of the solutions
at a modest computational cost. Finally, we summarize our conclusions
in Sec. IV.  The Appendix contains an outline of the derivation of
the Coulomb matrix element with these basis functions.  The analytical
expression presented there greatly simplifies our calculations.

\section{ Theoretical Method }

For concreteness, and to simulate recent quantum dot
systems,\cite{12,13,14} we assume quantum dots with an oblate
spheroidal profile where the lateral $xy$ confinement is much weaker
(or larger size) than that along the $z$ direction.  Correspondingly,
the electrons and holes are assumed confined in an effectively
two-dimensional potential with a constant $z$ profile, $V_z$.  We
assume $V_z$ to be a hard-wall confinement potential, so that the
$z$ component of the energy is $\hbar^2 \pi^2/2mL_z^2$, with $L_x, L_y
\gg L_z$.  Further, we approximate the single $z$ wavefunction in the
problem as a $\delta$ function centered at the origin, so that the problem
can be described by a separable Hamiltonian in two dimensions.  The
lateral confinement is modeled via harmonic potentials with two
different frequencies $\omega_x$ and $\omega_y$, which yield the
elliptical cross section of the dots with axes ratio given by $L_x /
L_y = \sqrt{\omega_y/\omega_x}$, for both electrons and holes.   The
smoothly varying potential should mimic well the situation in
experiments  where the dots are effectively embedded in a dielectric
matrix.\cite{1,12}

The effective-mass parabolic-band Hamiltonian for an electron-hole pair
is given by $H = H_e + H_h + H_{e-h}$, where the subscripts $e$ and $h$
represent electron and hole, and 
 \begin{equation}
H_e = \frac{p^2}{2m_e} +  \frac{1}{2}
m_e \omega_x^2 x_e^2 + \frac{1}{2} m_e \omega_y^2 y_e^2 + V_{ze} \, ,
 \end{equation}
 with a similar expression for the Hamiltonian of the hole,
$H_h$.\cite{NOTE}  The Coulomb interaction between electron and hole is
screened by a background dielectric constant $\epsilon$, so that
$H_{e-h} = -e^2/\epsilon r_{e-h}$.  
                                               
We rewrite the Hamiltonian into relative and center of mass
coordinates, described by ${\bf r} = {\bf r}_e - {\bf r}_h$, and
${\bf R} = (m{_e}{\bf r}_e + m_h{\bf r}_h)/M$.
 The total and reduced masses are given by $M = m{_e} + m{_h}$, and
$\mu = m{_e}m{_h}/M$, respectively.  The total Hamiltonian of this
system can then be written in the form 
$H = H{_{c.m.}} + H{_{rel}}$,
with the expected expressions  
 \begin{equation}
H_{c.m.} = \frac{P^2}{2M} + \frac{1}{2}M\omega_x^2
X^2  + \frac{1}{2}M\omega_y^2 Y^2 + V_Z \, ,
 \end{equation}
 and 
 \begin{equation}
H{_{rel}} = \frac{p{{^2}}}{2\mu} + \frac{1}{2}{\mu}\omega_x^2 x{^2}
 + \frac{1}{2}{\mu}\omega_y^2 y{^2}
 - \frac{e{^2}}{\epsilon \sqrt{x{^2} + y{^2}}} + V_z \, .
 \end{equation} 
 The Hamiltonian of the center of mass is obviously a two-dimensional
harmonic oscillator in the $XY$ plane, with wavefunction  $\Psi_{N_X
N_Y} = \phi_{N_X}(X)\, \phi_{N_Y}(Y)$, and energy $E_{c.m.}$, where
 \begin{equation} 
 \phi{_{N}}(X)  = \left(\frac{{\alpha_M}}{{\pi^{1/2}}2{^{N}}N!}\right)^{1/2} \,
e^{-\alpha_M^2 X^2/2} H_N \left( \alpha_M X \right) \, , 
 \end{equation} 
$\alpha_M = \sqrt{M \omega_x / \hbar} \, ,$ and 
 \begin{equation}
E_{c.m.} = \left(
N_X + \frac{1}{2} \right) \hbar \omega_x + \left( N_Y + \frac{1}{2} \right)
\hbar \omega_y + \frac{\hbar^2 \pi^2}{2ML_z^2} \, .
 \end{equation}
  Here, $N{_X}$ and $N{_Y}$ are quantum numbers for the center of mass
coordinate, and $H{_N}$ is a Hermite polynomial.\cite{17}

The physics of the problem is determined to a great extent by the ratio
between the effective Bohr radius, $a_B^* = \hbar^2 \epsilon / \mu
e^2$, and the size of the dot, $L=\sqrt{L_x L_y}$, where
$L_i=\sqrt{\hbar /\mu \omega_i}$.  The strong confinement limit for $L
\leq a_B^*$ is characterized by a weak electron-hole correlation and by
the Coulomb term being a small perturbation of the single-particle
confined-level energy.  On the other hand, the weak-confinement limit
for $L \geq a_B^*$ reduces asymptotically to the problem of a free
two-dimensional exciton for large $L$, where the Coulomb interaction 
dominates the state of the exciton \cite{1}.   
                                          
With this in mind, the effects of the Coulomb term $H_1$ in $H_{rel} =
H_0 + H_1$, are treated by using the solutions of $H_0$ as the basis
set in the diagonalization of $H_{rel}$. The unperturbed Hamiltonian of
the relative coordinate $H_0$ is also a two-dimensional harmonic
oscillator, so that the wavefunction of the interacting electron-hole
pair is described by a linear combination of wavefunctions,  $\psi_{n_x
n_y}=\phi_{n_x}(x) \phi_{n_y}(y)$, with the $\phi$'s  satisfying a
similar expression to Eq.\ (4), and correspondingly  
 \begin{equation}
E{_{rel}^0} =
\left( n_x + \frac{1}{2} \right) \hbar \omega_x + \left( n_y +
\frac{1}{2} \right) \hbar \omega_y + \frac{\hbar^2 \pi^2}{2\mu (2L_z)^2} \, ,
 \end{equation} 
 where $n_x$ and $n_y$ are quantum numbers for the relative coordinate.
(Notice $z$ confinement length for this coordinate is $2L_z$.)
 
With this basis set, the interaction matrix elements of the
electron-hole pair can be calculated analytically and expressed in
terms of hypergeometric functions as outlined in the Appendix.  This
analytical expression greatly simplifies the calculation, as most of
the computational time is spent on the calculation of the matrix
elements rather than on the diagonalization of the matrix.   The
resulting Hamiltonian matrix is real, symmetric, and sparse.  The
energies and eigenfunctions are calculated from the numerical
diagonalization of the matrix, for a given size of the basis.  The
diagonalization is repeated with larger basis sets until the desired
convergence is achieved (see below).
 
\subsection {Circular limit}  
 As an additional test of our numerical procedures we use the resulting
radial equation of the relative Hamiltonian for the circular dot case,
$\omega_x = \omega_y = \omega$, which is then given by 
 \begin{equation}
 H_{rel} = \frac{p^2}{2\mu} + \frac{1}{2} \mu \omega^2 r^2 - 
 \frac{e^2}{\epsilon r} \, .
 \label{neweq7}
 \end{equation}
 The resulting one-dimensional equation can be directly integrated
numerically, as reported by Que.\cite{7}  Furthermore, the problem can
also be solved using a harmonic basis set of radial states as those
described above.  For a large enough basis set, this approach yields
the same results as those obtained by direct numerical
integration.\cite{7}  Notice also that the large dot limit $(\omega
\approx 1/L \approx 0)$ is easily solved as an expansion in terms of
the Laguerre polynomial-based solutions of the free exciton
problem.\cite{7}  This allows one to obtain very accurate solutions for
$1/L \approx 0$ with little computation.  These results, moreover,
allow us to test the convergence of the irreducible two-dimensional
nonsymmetric problem $(\omega_x \neq \omega_y)$, as we discuss below.
                                          
\subsection {Optimized basis}
  Solutions for these elliptic cylinderlike quantum dots are also
carried out using an optimized basis set, as the solution method
discussed above requires a rather large number of states for full
convergence, especially for large dots.  Here, one notices that for
large dots it is the Coulomb interaction that dominates the exciton
states (since confinement becomes less important).  Correspondingly, we
choose a set of optimized frequencies, ${\Omega}_x$ and ${\Omega}_y$, 
which are larger than the original frequencies,  ${\omega}_x$ and
${\omega}_y$.   The values of the $\Omega$'s are determined
variationally, and allow one to consider $H$-matrix systems that are
much smaller than those required when one uses the $\omega$ basis.  
The physical reason for this is that as the dot size increases, the
exciton size converges to $a_B^{2D}$ (the radius of the free
two-dimensional exciton), and one needs a large number of
$\omega$ states to describe the small-scale structure of the exciton.

Notice that the harmonic $\Omega$ basis allows also an easy calculation
of  the $H$-matrix elements in this case, so that for example,
 \begin{eqnarray}
  \langle n_x{^{\prime}} \, n_y{^{\prime}} \, &| &H_{0} | n_x \, n_y
\, \rangle_\Omega =  \left[{\hbar}\Omega{_x}(n{_x}+\frac{1}{2}) \, + \,
{\hbar}\Omega{_y}(n{_y}+\frac{1}{2}) \right. \nonumber  \\ \,
 & & - \, \left.
\frac{\hbar}{2\Omega{_x}}(\Omega_x^2-\omega_x^2)(n{_x}+\frac{1}{2})
\, - \,
\frac{\hbar}{2\Omega{_y}}(\Omega_y^2-\omega_y^2)(n{_y}+\frac{1}{2})
\right] \, \delta{_{n{_{x^\prime}},n{_x}}} \,
\delta_{n_{y^\prime},n_{y}} \nonumber \\ 
 & &- \,
\frac{\hbar}{4\Omega{_x}}(\Omega_x^2- \omega_x^2) \left[
\sqrt{n_x(n{_x}-1)} \, \delta{_{n{_{x^\prime}},n{_x}-2}} +
\sqrt{(n_x+1)(n{_x}+2)} \, \delta{_{n{_x^{\prime}},n{_x}+2}} \right]
 \, \delta{_{n_{y^\prime},n_y}} 
 \nonumber \\ 
 & & - \,
\frac{\hbar}{4\Omega{_y}}(\Omega_y^2-\omega_y^2) \left[
\sqrt{n_y(n{_y}-1)} \, \delta{_{n{_{y^\prime}},n_y-2}} +
\sqrt{(n_y+1)(n{_y}+2)} \, \delta{_{n{{_{y^{\prime}}}},n{_y}+2}}
\right] \, \delta{_{n{{_{x^\prime}}},n{_x}}}  \,\,  .
 \label{oldeq7}
 \end{eqnarray}
 Notice this reduces to the obvious diagonal matrix for $\Omega
\rightarrow \omega$.  This expression allows one to evaluate the
Hamiltonian matrix rather conveniently, even for this other basis set.

\subsection {Exciton characteristics}             
 The wavefunctions of the relative coordinate problem can then be
written as $| \psi \rangle  = \sum_{n{_x}n{_y}}  a{_{n{_x}n{_y}}} |
n{_x} , n{_y} \rangle$, with either the $\omega$ or $\Omega$ basis
states, which can be used to study various characteristic properties of
the exciton system.  For example, the mean electron-hole separation
$r_s$, is given by
 \begin{eqnarray}
 r_s^2 &=& \langle \psi | r^2 | \psi \rangle  =  \sum_{n{_x}n_y}
 \left[\frac{\hbar}{{\mu}\Omega{_y}}
(n{_y}+\frac{1}{2})+\frac{\hbar}{{\mu}\Omega{_x}}(n{_x}+\frac{1}{2})\right]
|a{_{n{_x},n{_y}}}|^2 \, \nonumber \\ 
 & &+
 \, \frac{1}{2}\frac{\hbar}{{\mu}\Omega{_y}}\left[\sqrt{(n{_y}+2)(n{_y}+1)}
 \, a{{^*}{_{n{_x},n{_y}+2}}} +
\sqrt{n{_y}(n{_y}-1)}
 \, a{{^*}{_{n{_x},n{_y}-2}}}\right]a{_{n{{_x}},n{{_y}}}} \nonumber \\
 & &+ \,
\frac{1}{2}\frac{\hbar}{{\mu}\Omega{_x}}\left[\sqrt{(n{_x}+2)(n{_x}+1)}
  \, a{{^*}{_{n{_x}+2,n{_y}}}} +
\sqrt{n{_x}(n{_x}-1)}
  \, a{{^*}{_{n{_x}-2,n{_y}}}}\right]a{_{n{{_x}},n{{_y}}}} \, ,
 \end{eqnarray}
 which gives an idea of the exciton size.

 One can also use the diagonalization results to calculate
directly measurable properties, such as the linear optical
susceptibility of the quantum dot/disk.  The linear optical
susceptibility is proportional to the dipole matrix elements between
one electron-hole pair j state and the vacuum state, $\langle  0 | P |1 
\rangle_j$.\cite{5}  These in turn are proportional to the interband
matrix element, $p_{cv}$,\cite{18} which is the matrix element formed
between an electron and hole in the conduction and valence bands,
respectively.  The form of the dipole matrix elements for a single
exciton in the envelope function approximation is given by\cite{5,7}  
 \begin{equation}
|\langle 0 | P |1 
\rangle|^2 = |p_{cv}|^2 {|{\psi}(0)|}^2 \left|{\int}{\int} {\Psi}(X_e,Y_e)
 \, dX_e \, dY_e \right|^2 \, .
 \end{equation}
  Here, the wavefunction for the relative coordinate is given as
above, so that 
 \begin{equation}
 |\psi(0)|^2 = (\mu/\hbar \pi) \sqrt{w_ x w_y} \, \left|\sum_{n_xn_y}
(2^{n_x+n_y}n_x!n_y!)^{-1/2} \, a_{n_xn_y}\right|^2 \, ,
 \end{equation}
 where the $ a_{n_xn_y}$ coefficients are obtained from the diagonalization
of the relative-coordinate Hamiltonian, and
 \begin{equation}
 \left|{\int}{\int} {\Psi}(X_e,Y_e) dX_e dY_e \right|^2 = 4 \pi^2 \hbar
N_X! \, N_Y! \, \left[\pi M \sqrt{w_xw_y} \,\, 2^{N_X+N_Y} \,
(N_X/2)!^2 \, (N_Y/2)!^2 \right]^{-1} \, , 
 \end{equation}
 with $N_X= {\rm even}$ and $N_Y = {\rm even}$, for nonzero matrix
elements.  Finally, the dipole matrix elements have the form  
 \begin{equation} 
 |\langle0| P |1\rangle|^2 = 4 |p_{cv}|^2 \frac{\mu}{M} N_X! \, N_Y! \,
\left[2^{N_X+N_Y} \, (N_X/2)!^2 \, (N_Y/2)!^2 \, \right]^{-1} \, \left|
\sum_{n_x n_y}  (2^{n_x+n_y}n_x! \, n_y!)^{-1/2} \, a_{n_x n_y} \,
\right|^2 \, .
 \end{equation}
 The linear optical susceptibility can then be calculated from
 \begin{equation}
\chi ( \omega ) = \sum_{j} |\langle 0 | P | 1 \rangle_j |^2 (\hbar
\omega -E_j -i \hbar \Gamma)^{-1} \, ,
 \end{equation}
 where $\Gamma$ is introduced as a phenomenological level broadening
constant.
  
Notice also that since experimental systems are typically configured
to analyze a large collection of nearby dots, one should in principle
be concerned by the effect of local fields.   However, in typical
experimental systems so far, where the separation between dots can be
several microns, it is valid to assume that dots are basically
independent.  In the case of higher-dot densities, however, the
dynamical response of the system may be affected by the local fields
produced by neighboring dots, and one can obtain that response from the
individual microscopic polarizabilities.\cite{16}

\section { Results }

As an interesting example of a typical system, we use parameters to
describe GaAs quantum dots, so that the dielectric constant is
$\epsilon=13.1$, and the carrier masses are $m{_e}=0.067m{_0}$, and (the
heavy hole effective mass) $m_{hh}=0.37m{_0}$.\cite{NOTE}  We present
the numerical results for heavy-hole excitons in GaAs quantum dots with
elliptical and circular cross sections.  The solutions can be
calculated using a sufficiently large basis set and/or an optimized
basis set, as described above.  Results obtained from these methods and
different basis sets are shown in Fig.\ 1.  The exciton binding energy
and normalized electron-hole separation are shown as a function of
quantum dot size ranging from 2 to 100 nm.
                     
In the insets, results are shown for {\em circular} dots, with dot,
dashed, and dot-dashed curves showing results for basis sets with
$M=30, 100$, and $500$ wavefunctions, respectively.  Here, states with
$n$ and $n{^{\prime}}$ from $0$ to $29$, $99$, and $499$ are used in
Eq.\ (\ref{neweq7}).  (The matrix size is obviously $M \times M$, and
is diagonalized by a QL decomposition technique.\cite{19})  The results
of the one-dimensional radial equation in the weak-confinement limit
are shown with the solid line for comparison, and represent the exact
quantity (both $E_b$ and $r_s$) for large $L$. The transition between
the strong and weak confinement regimes comes appropriately when the
size of the quantum dot is near the effective Bohr radius, $L \approx
a_B^* =12.2$ nm.  Notice that it is for $L \approx 15$ nm that the
$M=100$ curve (dashed) departs from the exact result (solid line), and
that one requires larger $M$ values as $L$ increases to achieve better
convergence.  The $M=500$ basis set  (dot-dashed curve) yields the
convergent solutions with acceptable accuracy and execution time for a
larger range of $L$ values ($\leq 60$ nm).  Similar behavior can be
seen in the electron-hole separation in the ground state [inset in
panel (b)].

The inset in 1(a) also shows the difference between the $\omega$ basis
and $\Omega$ basis  results.   Diamonds show results for the $M=400$
basis set with optimized $\Omega$ frequency, while triangles show
results for $M=400$ with the original $\omega$ basis set.   For both
cases, the states with $n_x$, $n_y$, $n_x{^{\prime}}$, and
$n_y{^{\prime}}$ from $0$ to $19$, respectively, are included for the
$M=400$  basis set (see Eq.\ \ref{oldeq7}).  These states are used as
the basis set for elliptical quantum dots.  The results with the
optimized basis set are essentially identical to the convergent
solutions.   The optimized frequencies used are also given as a
function of dot size in the  inset 1(a) (with $E_\Omega = \hbar
\Omega$) as a long-dashed curve.  For dot sizes $30$ nm and larger, the
optimized frequency $\Omega$ converges to 80 meV, corresponding to a
dot size $L_{\Omega}=\sqrt{\hbar/{\mu}{\Omega}} =4.1$ nm.  This latter
value is close to the two-dimensional effective Bohr radius
($a^{2D}_B=a^*_B/2\sqrt{2}=4.3$ nm), as one would expect.  Notice that
for most of the range of $L$'s shown, one is in the weak-confinement
regime, where $L \ge a^{2D}_B$.   This is why the optimized
basis set gives very reasonable results with minimal effort.  It is
also interesting that the agreement continues also for smaller dot
sizes, completing the range from $2$ to $100$ nm.  The exciton ground
state binding energies and the normalized electron-hole  separation
obtained with the optimized $\Omega$ basis approach are basically exact
to the fully converged results.

The main panels in Fig.\ 1 (a) and (b) show the geometrical confinement
effects of excitons with several different size axis ratios in the
$xy$ plane.  The plots show results versus $L=\sqrt{L_xL_y}$, the
effective size of the dot, for ${\omega_x}$/${\omega_y}=1$, $4$, and
$9$ ($L_y/L_x=1$, $2$, and $3$) with diamonds, pluses, and triangles,
respectively.  
The exciton binding energy increases for a small $L$ as the axis 
ratio is increased; meanwhile, the normalized electron-hole separation 
is basically unchanged, 
except for small $L$ values, where
the confinement energy dominates.  Notice that as the axis ratio
increases, the single-particle and exciton states move up in energy but 
the {\em binding} energy increases.  
This increase in binding energy for
the elliptical dots is then related to the increase of the Coulomb
energy {\em relative} to the confinement contribution to the total
energy.  (In fact, since $r_s/L$ is basically unchanged with geometry,
the Coulomb interaction energy is nearly constant in all these cases.)

To better explore the geometrical confinement effects on excitons,  we
show the linear optical susceptibility of the GaAs quantum disks with
elliptical cross sections and lateral mean size of $L=\sqrt{L_xL_y}=5$
nm (Fig.\ 2), and 10 nm (Fig.\ 3).  The dot thickness along the $z$-
direction is kept constant at $L_z=3$ nm, and several size ratios for
each axis in the $xy$ plane are shown.  We use here also a value for
the optical bandgap of $E_g = 1.51$ eV.  The results presented here
were obtained using the optimized $\Omega$ basis set approach discussed
above.  Notice that since this function represents all of the possible
transitions of this excitonic system, its features would be measurable
via photoluminescence excitation measurements.  On the other hand,
the photoluminescence response would correspond to the first (lower
energy) feature in these traces, associated with the ground state of
the excitonic system.

Figure 2 shows the imaginary part of the linear optical susceptibility
as a function of frequency for a dot with $L = \sqrt{L_x L_y} = 5$ nm
(a broadening of $\Gamma=2$ meV is used).  The bottom trace is for a
circular dot, so that $L_x = L_y = 5$ nm.  The upper two traces show
results for elliptical dots with a size ratio  $L_y:L_x = 2:1$
($L_y=7.0$ nm, and $L_x=3.5$ nm), and $3:1$ ($L_y=8.7$ nm, $L_x=2.9$
nm), respectively, all having the same mean size $L=5$ nm.  [Notice
that  although the peak heights are in arbitrary units, the ratio
between different peaks or traces is real, reflecting the different
dipole matrix elements involved.]~  One obvious difference among traces
is that the first transition energy shifts to higher values by $\approx
26$ and 69 meV, as the size ratio changes from 1:1 to 2:1 and 3:1,
respectively.  Notice that while increasing the size ratio, the exciton
binding energies {\em increase} from 47 meV to 48 and 49 meV for each
ratio.  It is then clear that the larger blue shifts are due mostly to
the increasing confinement as the disk becomes more elliptical, and not
due to the binding energy between electron and hole.  In
all the cases shown, the transition involving the ground state of the
exciton is dominant, as the excited states appear only with smaller
oscillator strength.  Notice further that for the larger length ratio,
the spectrum is understandably sparser, as the levels associated with
the narrow dimension are quickly pushed upwards in energy.  Finally,
since the elliptical dots have lower symmetry, accidental degeneracies
are fewer and the transition peaks show nearly monotonically decreasing
intensities (unlike the circular dot case). [Notice these all the
susceptibility traces include transitions between different
center-of-mass states up to the $N_X=N_Y=10$ levels.  Additional $N_X$
and $N_Y$ values would yield higher energy structure.]

Figure 3 shows also the imaginary part of the susceptibility but for a
dot size of 10 nm, and here with level broadening of 1 meV.\@ The first
transition energy in the circular quantum dot is 1.78 eV, while the
values in the elliptical dots are 1.79 and 1.80 eV, for ratios $2:1$
and $3:1$.  The transition energies result here to be lower than in
Fig.\ 2  since the confinement is not as strong, reducing the effective
gap energy.   In this set of curves, the first transition energy
(involving the ground state of the exciton) is shifted upwards for
elliptical dots, due to
the increased confinement, although the shift is not as large as in
Fig.\ 2, since the overall lengths are larger.  The first excited state
appears split because the two-fold degeneracy of the excited state is
broken as the dot becomes elliptical.

As dot size is increased, it is apparent that the geometrical effects
are not as prominent, producing only a small shift of the spectrum of
transitions.  Incidentally, the onset of transitions for larger size
dots ($L \approx 100$ nm) compares qualitatively well with experimental
photoluminescence spectra in dots with similar disk geometry,\cite{10}
where features appear in the energy range 1.73 -- 1.74 eV, for dots
with radius thought to be in the range 150--200 nm.  According to the
experimental results, the additional confinements give the observed
blue shift, compared to two-dimensional quantum-well exciton case.
Similarly, a blue shift appears due to the elliptical shape, although
for $L \approx 100$ nm they are only $\leq 10$ meV, as the size ratio
increases to 2:1 and 3:1.

As an example of the effects for different materials, Fig.\ 4 shows the
imaginary part of the susceptibility of InAs quantum dots with lateral
mean size of $12$ nm and thickness of $2.8$ nm --- having a dielectric
constant $\epsilon=14.6$, $m{_e}=0.026m{_0}$, and heavy hole effective
mass $m_{hh}=0.41m{_0}$.  Here, the energy gap is taken as 0.43 eV, and
we also use the optimized $\Omega$ basis set to obtain the results
shown.  The convergent optimized size, $L_{\Omega}=10.5$ nm, is close
to the two dimensional effective Bohr radius ($a^{2D}_B=10.54$ nm).  In
this case, the first transition energy shifts to higher values by
$\approx 11$ and 28 meV, as the size ratio changes from 1:1 to 2:1 and
3:1, respectively.  Notice the similarity with Fig.\ 2, although the
energy scale and size (12 nm) are completely different here. This
similarity is due to the scaling of the problem in terms of $a^*_B$.
For these InAs parameters, $a^*_B=29.8$ nm, so that $a^*_B/L \approx
2.5$, comparable to the value in Fig.\ 2 for GaAs, where $a^*_B=12.2$,
$L=5$, and $a^*_B/L \approx  2.4$.

\section{ Conclusions}
 
We have demonstrated that strong geometrical confinement effects appear
on excitons in GaAs and InAs quantum dots with elliptical
cross sections. The solutions have been obtained using sufficiently
large basis sets as well as with an optimized basis set. The results
obtained with the optimized basis sets, such as exciton binding energy
and normalized electron-hole separation, are extremely close to the
best converged results, and only with a relatively modest computational
effort.

The linear optical susceptibilities are calculated for several
different lateral size ratios of each axis ($x$ and $y$). Strong blue
shifts in the susceptibilities are observed as the size ratio is
increased and the shifts due to the geometrical shape effects are
especially important for the smaller dot sizes ($ \leq 25$ nm).  The
shifts are due mostly to the increasing confinement as the dot becomes
more elliptical, and not due to the interaction energy between
electron and hole.  A splitting of the first few excited states appears
in the elliptical cross section cases since the symmetry-related
degeneracy of the excited states in the circular dot is broken.  This
gives also rise to a more monotonic decrease of the peak intensities
seen as the energy of the transition increases.

   \acknowledgments

We would like to thank R.L. Cappelletti and D.A. Drabold for helpful
discussions, and the support of the US Department of Energy through
grant no.\ DE--FG02--91ER45334.  Calculations were partially performed
at the Cray Y/MP of the Ohio Supercomputer Center.  S.E.U. acknowledges
support of the A. v. Humboldt Foundation.

 \appendix                                                   
                             
\section*{}  

The Coulomb interaction matrix elements of the electron-hole pair can
be calculated analytically using the harmonic-oscillator basis sets we
use.  With the basis set in Eq.\ (4), one can write
 \begin{eqnarray}
 \langle n{{_x}^{\prime}},n{{_y}^{\prime}} |
\frac{e^2}{\epsilon\sqrt{x^2+y^2}} &|& n_x,n_y \rangle  =
\frac{e^2}{\epsilon\pi} \sqrt{\frac{\mu \omega_x}{\hbar}} \,
\sqrt{\frac{\mu \omega_y}{\hbar}} \, (2^{n_x^{\prime} + n_x + n_y^{\prime}
+ n_y} \,
n_x^\prime! \, n_x! \, n_y^\prime! \, n_y!)^{-1/2} \nonumber \\
 & & 
 \times \, \int_{-\infty}^{\infty} \int_{-\infty}^{\infty} \, dx \, dy \,
e^{-(\frac{\mu\omega_x}{\hbar}x^2 + \frac{\mu\omega_y}{\hbar}y^2)} \,
\frac{1}{\sqrt{x^2+y^2}} \, \nonumber \\
& & \times H_{n_{x^{\prime}}}(\sqrt{\frac{\mu\omega_x}{\hbar}}
x) \, H_{n_{x}}(\sqrt{\frac{\mu\omega_x}{\hbar}} x) \,
 H_{n_{y^{\prime}}}(\sqrt{\frac{\mu\omega_y}{\hbar}} y) \,
 H_{n_{y}}(\sqrt{\frac{\mu\omega_y}{\hbar}} y) \, .
\end{eqnarray}   
As the Hermit polynomials are represented by,\cite{17}
\begin{equation} H_{n_{x}}(\sqrt{\frac{\mu\omega_x}{\hbar}} x) \, = \,
\sum_{\beta=0}^{n_x/2} \, (-1)^{\beta} \, 2^{n_x-
2\beta}(\frac{\mu\omega_x}{\hbar})^{\frac{n_x-2\beta}{2}}\frac{n_x!}{\beta! \,
(n_x-2\beta)!} \, x^{n_x-2\beta} \, , \end{equation}
one can write,
 \begin{eqnarray}
 \langle n{{_x}^{\prime}},n{{_y}^{\prime}} |
\frac{e^2}{\epsilon\sqrt{x^2+y^2}} &|& n_x,n_y \rangle =
\frac{e^2}{\epsilon\pi} \sqrt{\frac{\mu \omega_x}{\hbar}} \,
\sqrt{\frac{\mu \omega_y}{\hbar}} \, (2^{n_x^{\prime} + n_x + n_y^{\prime}
+ n_y} \,
n_x^\prime! \, n_x! \, n_y^\prime! \, n_y!)^{-1/2} \nonumber \\ \nonumber \\
 & & \times \, \sum_{\alpha=0}^{[n_x^\prime /2]} \sum_{\beta=0}^{[n{_x}/2]}
\sum_{\gamma=0}^{[n_y^\prime /2]} \sum_{\delta=0}^{[n_y /2]}
\frac{(-1)^{\alpha+\beta+\gamma+\delta} \,
n_x^\prime! \, n_x! \, n_y^\prime! \, n_y!}{\alpha!
(n_x^\prime -2\alpha)! \beta! (n_x -2\beta)! \gamma! (n_y^\prime
-2\gamma)! \delta! (n_y -2\delta)!} \nonumber \\ \nonumber \\
 & & \times \, (\frac{\mu\omega_x}{\hbar})^{\frac{n_{x^{\prime}}+n_x-2\alpha-
2\beta}{2}} \, (\frac{\mu\omega_y}{\hbar})^{\frac{n_{y^{\prime}}+n_y-2\gamma-
2\delta}{2}} \, 2^{n_{x^{\prime}}+n_x+n_{y^{\prime}}+n_y-2\alpha
-2\beta-2\gamma-2\delta} \nonumber \\ \nonumber \\
 & & \times \, \int_{-\infty}^{\infty} \int_{-\infty}^{\infty}
\, dx \, dy \,
e^{-(\frac{\mu\omega_x}{\hbar}x^2 + \frac{\mu\omega_y}{\hbar}y^2)} \,
\frac{1}{\sqrt{x^2+y^2}} \, x^{n_{x^{\prime}}+n_x-2\alpha-
2\beta} \, y^{n_{y^{\prime}}+n_y-2\gamma-
2\delta} \, .   \end{eqnarray}  
 Here, the integral factor,
\[ I = \int_{-\infty}^{\infty} \int_{-\infty}^{\infty}
\, dx \, dy \, e^{-(\frac{\mu\omega_x}{\hbar}x^2 +
\frac{\mu\omega_y}{\hbar}y^2)} \,
\frac{1}{\sqrt{x^2+y^2}} \, x^{n_{x^{\prime}}+n_x-2\alpha-
2\beta} \, y^{n_{y^{\prime}}+n_y-2\gamma-
2\delta} \, , \]
 can be transformed to polar coordinates, so that, 
\begin{eqnarray}
 I \, &=& \, \int_{0}^{2\pi} \, d{\varphi} \, \left[ \int_{0}^{\infty} \,
dr \, e^{-(ar^2\cos^2{\varphi}+br^2\sin^2\varphi)} \, r^{l+m}\right]
 \, \cos^l{\varphi} \, \sin^m{\varphi} \nonumber \\
 &=&\, (-1)^{2l+m} \, \Gamma(n) \, \frac{1}{a^n} \,
\frac{\Gamma(\frac{l+1}{2}) \, \Gamma(\frac{m+1}{2})}{\Gamma(\frac{l+m}{2}+1)}
 \, F(\frac{m+1}{2}\, ,\, n \, ;\, \frac{l+m}{2}+1 \, ; \, \frac{a-b}{a}) \, ,
 \end{eqnarray}
where $a=\mu\omega_x / \hbar$, $b= \mu\omega_y / \hbar$,
$l=n_{x^{\prime}}+n_x-2\alpha-2\beta$, $m=n_{y^{\prime}}+n_y-2\gamma-
2\delta$, and $n=\frac{1}{2} (l+m+1)$.
  
The interaction matrix elements can be expressed in terms of
hypergeometric functions as
 \[ \langle n{{_x}^{\prime}},n{{_y}^{\prime}} |
\frac{e^2}{\epsilon\sqrt{x^2+y^2}} | n_x,n_y \rangle =
\frac{e^2}{\epsilon\pi} \sqrt{\frac{\mu \omega_y}{\hbar}} \, (2^{s_x + s_y} \,
n_x^\prime! \, n_x! \, n_y^\prime! \, n_y!)^{-1/2} \]
 \[ \times \sum_{\alpha=0}^{[n_x^\prime /2]} \sum_{\beta=0}^{[n{_x}/2]}
\sum_{\gamma=0}^{[n_y^\prime /2]} \sum_{\delta=0}^{[n_y /2]}
\frac{(-1)^{\eta} \, n_x^\prime! \, n_x! \, n_y^\prime! \, n_y!}{\alpha!
(n_x^\prime -2\alpha)! \beta! (n_x -2\beta)! \gamma! (n_y^\prime
-2\gamma)! \delta! (n_y -2\delta)!} \]
 \[ \times \left( \frac{\omega_y}{\omega_x} \right)^{\frac{s_y}{2} -\gamma
-\delta} (-1)^{s_y} \, 2^{s_x +s_y -2\eta} \]
 \[ \times F \left[ \frac{1}{2} (s_y +1) -\gamma -\delta,  \frac{1}{2}
(s_x +s_y +1) -\eta ; \frac{1}{2} (s_x +s_y) -\eta +1; 1
-\frac{\omega_y}{\omega_x} \right] \]
\begin{equation}
 \times \, \Gamma \left[ \frac{1}{2} (s_x +s_y +1)  -\eta \right] \,
\Gamma \left[ \frac{1}{2} (s_x +1) -\alpha -\beta \right]
\Gamma \left[ \frac{1}{2} (s_y +1)  -\gamma -\delta \right]
/\Gamma \left[ \frac{1}{2} (s_x +s_y)  -\eta +1 \right] \, , \end{equation}
 where $F$ is a hypergeometric function,\cite{17} $\eta = \alpha + \beta +
\gamma + \delta$, and the extra
constraints of $s_y= n_y +n_y^\prime = {\rm even}$, and $ s_x = n_x
+n_x^\prime = {\rm even}$, are required for this matrix element to be
non-zero.  Use of this equation in the calculation of the Coulomb matrix
elements was very important in the solution of the problem, as it reduces
the computation time substantially.

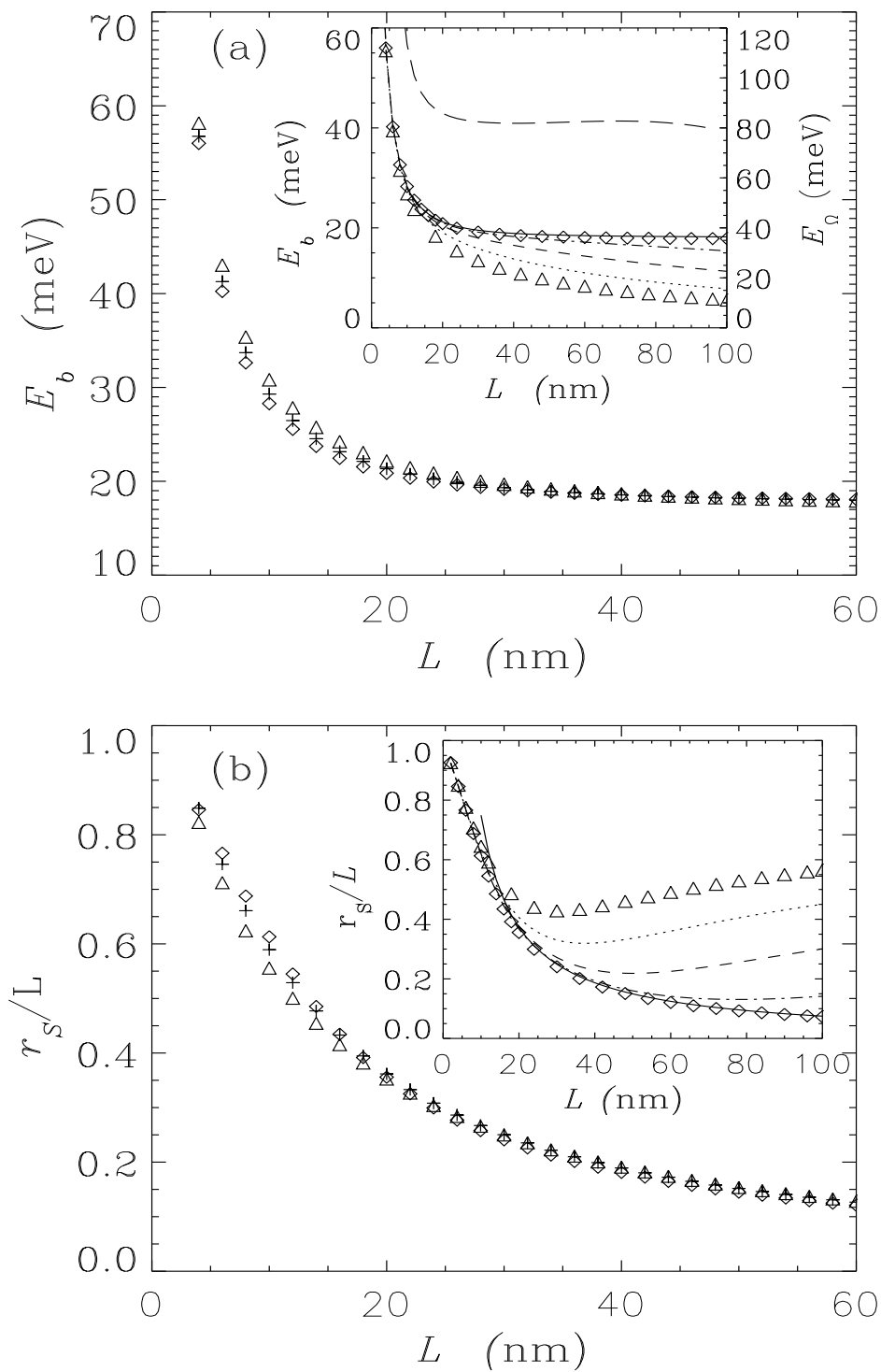
 \begin {figure}
  \caption{GaAs quantum dot.  (a) Exciton binding energy; (b) normalized
electron-hole separation as function of the quantum dot size for
elliptical quantum dots with several size ratios in the $xy$ plane.
$L_x:L_y$ shown are 1:1 to 2:1, and 3:1, with diamonds, pluses, and
triangles, respectively.  Insets: Solid, dot, dashed, and dashed-dot
curves are the results for circular case and $M=30$, 100, and 500 basis
sets, respectively.  Diamonds show results for the $M=400$
$\Omega$ basis set with optimized frequency, while triangles are for
$M=400$ in the original $\omega$ basis.   Inset in (a) also shows (scale
on the right) the value of $E_\Omega = \hbar \Omega$ used in the calculation. }
 \end   {figure}
  
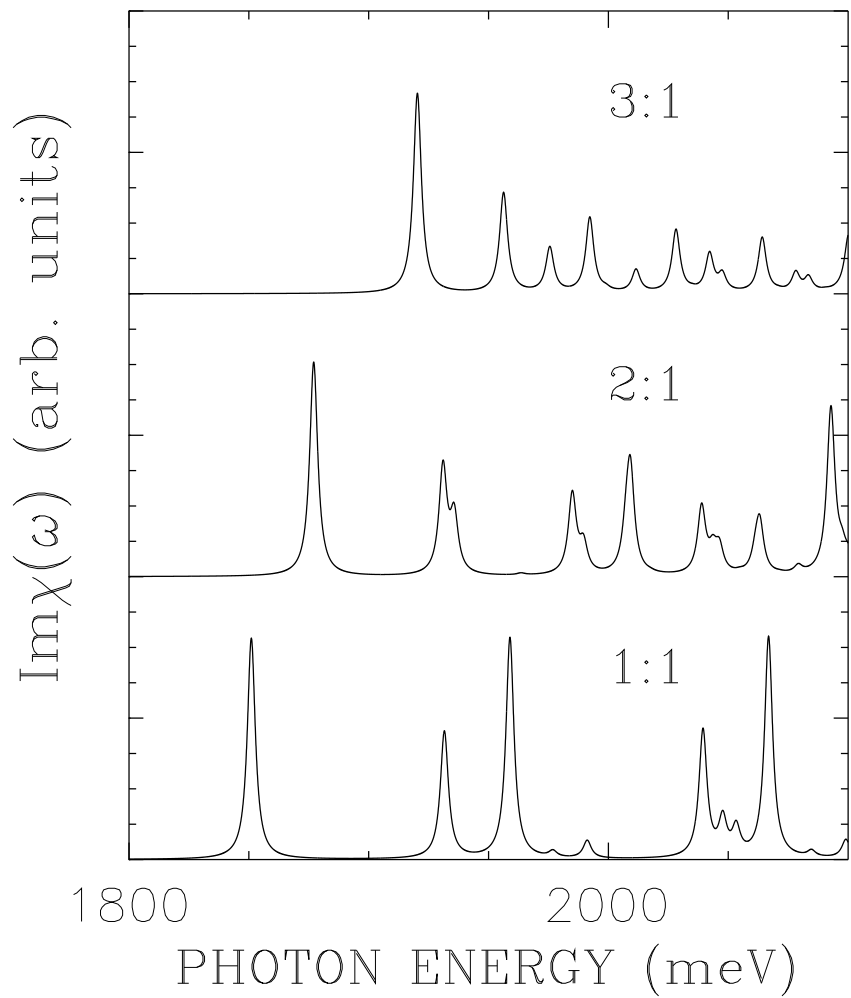
 \begin {figure}
   \caption{Imaginary part of linear optical susceptibility for a GaAs
elliptic quantum dot with the same area but different axis ratios, as a
function of frequency.  Bottom trace is for a circular dot, $L_x = L_y
= 5$ nm.  Upper two traces show results for elliptical dots with size
ratios  $L_y:L_x = 2:1$ ($L_y=7.0$ nm and $L_x=3.5$ nm), and 3:1
($L_y=8.7$ nm and $L_x=2.9$ nm), respectively.}
 \end {figure}

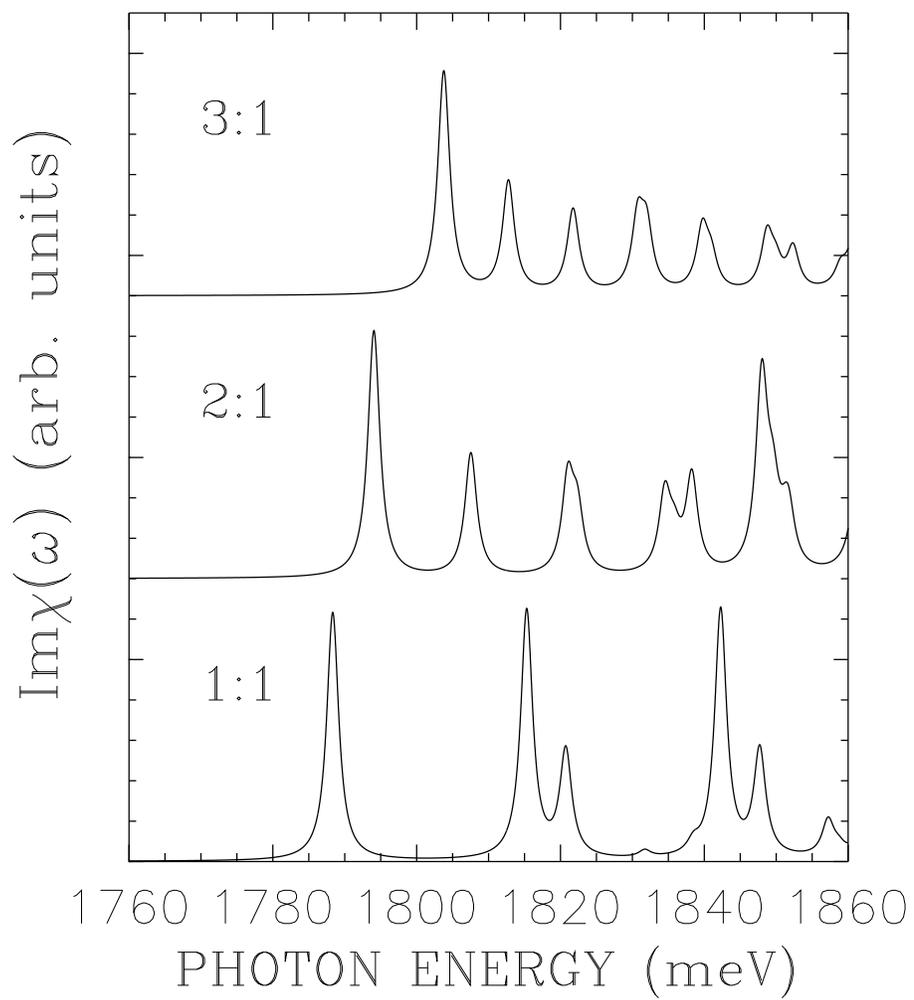
 \begin{figure}
   \caption{Same as Fig.\ 2 but for a size $L=10$ nm.}
 \end {figure}

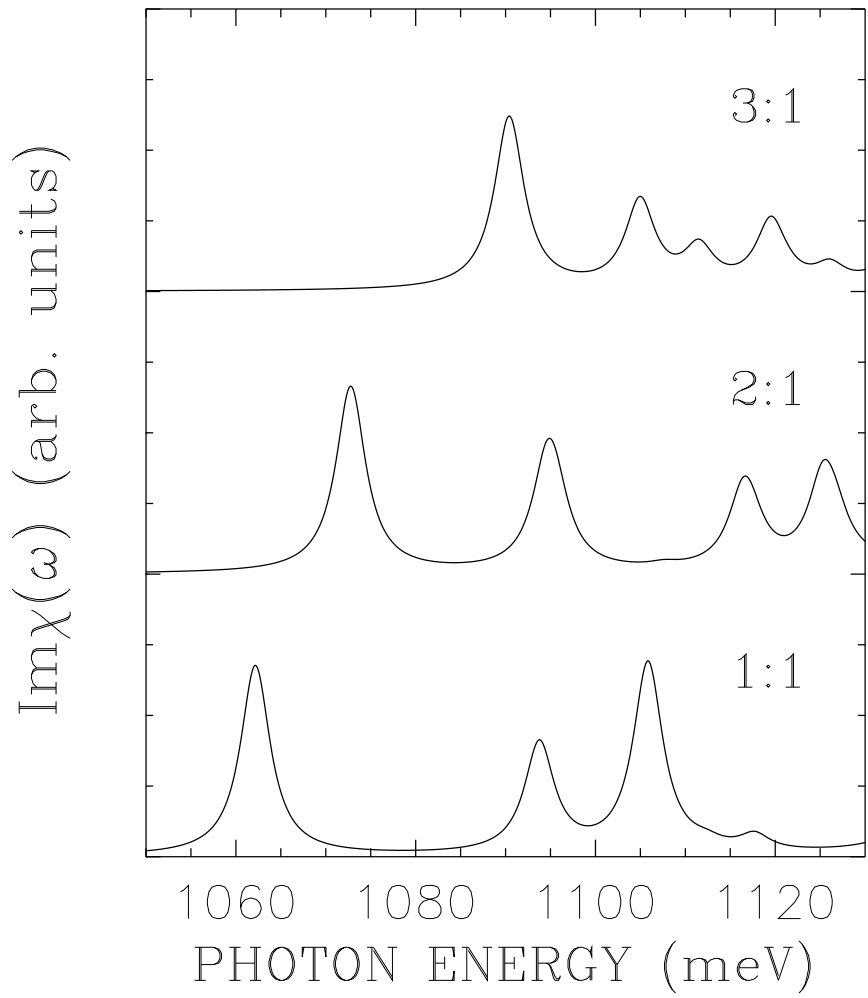
 \begin {figure}
   \caption{Imaginary part of linear optical susceptibility for InAs
elliptic quantum dots with lateral mean size $12$ nm and thickness
$2.8$ nm (with $\Gamma=2.0$ meV), as a function of frequency. The
bottom trace is for a circular dot with $L_x = L_y = 12$ nm.  The upper
two traces show results for elliptical dots with a size ratio  $L_y:L_x
= 2:1$ ($L_y=17$ nm and $L_x=8.5$ nm), and 3:1 ($L_y=20.7$ nm and
 $L_x=6.9$ nm), respectively.  Notice the similarity with Fig.\ 2.}
 \end {figure}

\end{document}